\documentclass[conference]{IEEEtran}
\IEEEoverridecommandlockouts
\usepackage{cite}
\usepackage{amsmath,amssymb,amsfonts}
\usepackage{algorithmic}
\usepackage{graphicx}
\usepackage{textcomp}
\usepackage{xcolor}

\usepackage{xcolor}
\usepackage{tikz}
\usepackage{balance}

\usepackage{pgfplots}
\usepackage{pgfplotstable}
\pgfplotsset{compat=1.7}
\usepgfplotslibrary{groupplots}
\usepackage{soul} %za komandu \hl
\sethlcolor{orange}
\usepackage[normalem]{ulem} % za komandu sout
\usepackage[scientific-notation=true]{siunitx}
\def\BibTeX{{\rm B\kern-.05em{\sc i\kern-.025em b}\kern-.08em
    T\kern-.1667em\lower.7ex\hbox{E}\kern-.125emX}}
\begin{document}
\definecolor{recondark}{RGB}{180,30,30}   % lambda = 0.7
\definecolor{reconlight}{RGB}{230,120,120} % lambda = 0.9
\definecolor{classdark}{RGB}{30,80,180}    % dark blue
\definecolor{classlight}{RGB}{120,170,230} % light blue

\title{%Semantic Split Learning: Bridging Task-Oriented Communication and Channel-Aware Inference
Task-Oriented Wireless Transmission of 3D Point Clouds: Geometric Versus Semantic Robustness\\
\thanks{This work is supported by the Serbian Ministry of Science, Technological Development and Innovation (Serbia–China Cooperation Project No. 00101957 2025 13440 003 000 620 021) and Intergovernmental International Science and Technology Innovation Cooperation of National Key Research \& Development Program of China under Grant 2024YFE0197400.}
}

\author{\IEEEauthorblockN{Vukan Ninkovic\IEEEauthorrefmark{1}\IEEEauthorrefmark{2}, Tamara Sobot\IEEEauthorrefmark{1}, Vladimir Vincan\IEEEauthorrefmark{1}, Gorana Gojic\IEEEauthorrefmark{1}, Dragisa Miskovic\IEEEauthorrefmark{1}, Dejan Vukobratovic\IEEEauthorrefmark{2} 
\vspace{1mm}
\IEEEauthorblockA{
\IEEEauthorblockA{\IEEEauthorrefmark{1}The Institute for Artificial Intelligence Research and Development of Serbia, Serbia
}
\IEEEauthorblockA{\IEEEauthorrefmark{2}Faculty of Technical Sciences, University of Novi Sad, Serbia}
}}}
\maketitle

\begin{abstract}
Wireless transmission of high-dimensional 3D point clouds (PCs) is increasingly required in industrial collaborative robotics systems. Conventional compression methods prioritize geometric fidelity, although many practical applications ultimately depend on reliable task-level inference rather than exact coordinate reconstruction. In this paper, we propose an end-to-end semantic communication framework for wireless 3D PC transmission and conduct a systematic study of the relationship between geometric reconstruction fidelity and semantic robustness under channel impairments. The proposed architecture jointly supports geometric recovery and object classification from a shared transmitted representation, enabling direct comparison between coordinate-level and task-level sensitivity to noise. Experimental evaluation on a real industrial dataset reveals a pronounced asymmetry: semantic inference remains stable across a broad signal-to-noise ratio (SNR) range even when geometric reconstruction quality degrades significantly. These results demonstrate that reliable task execution does not require high-fidelity geometric recovery and provide design insights for task-oriented wireless perception systems in bandwidth- and power-constrained industrial environments.
\end{abstract}

\begin{IEEEkeywords}
Semantic communication, task-oriented communication, 3D point clouds, industrial IoT.
\end{IEEEkeywords}

\section{Introduction}

High-resolution 3D perception has become a cornerstone of modern industrial automation, enabling collaborative robotics, inspection, and precision manufacturing. Depth and structured-light sensors provide detailed geometric representations of objects and workspaces in the form of 3D point clouds (PCs)~\cite{liu_2024}. In distributed industrial Internet of Things (IIoT) environments, these PCs are frequently transmitted over wireless links to support remote perception, monitoring, and decision-making. However, their high dimensionality, strict latency requirements, and limited wireless resources make efficient and robust transmission a fundamental challenge~\cite{Shao_2025}.

Conventional PC compression techniques primarily target geometric rate–distortion optimization. While effective for storage and visualization, such methods treat PCs as purely geometric signals and overlook the ultimate objective of many industrial systems: reliable task execution~\cite{cao20193d}. In practical scenarios, downstream processes, such as object recognition or manipulation, depend on semantic inference rather than exact coordinate-level reconstruction. This mismatch motivates a shift from signal-centric compression toward task-oriented communication design~\cite{saleh_2025}.

Semantic communication has recently emerged as a principled framework for transmitting task-relevant information rather than faithfully reconstructing signals~\cite{papas_2021}. For 3D PCs, prior works have primarily focused on either semantic-aware reconstruction~\cite{11361156,Liu_2025,saleh_2025} or robust task-oriented inference such as classification~\cite{10437861}. While these studies demonstrate the potential of semantic encoding under wireless impairments, they largely rely on generic benchmark datasets and treat reconstruction fidelity and semantic robustness as separate objectives.

In contrast, industrial IIoT scenarios, such as collaborative robotics, require both geometric consistency and reliable task-level decisions under realistic wireless constraints. The relationship between geometric distortion and semantic performance in such settings remains insufficiently understood. To address this gap, we develop an end-to-end semantic communication framework for wireless transmission of object-centric 3D PCs and evaluate it on a dedicated industrial collaborative robotics dataset. A unified dual-branch receiver performs both geometric reconstruction and semantic classification from the same transmitted latent representation, enabling a controlled empirical analysis of geometric and semantic robustness under additive white Gaussian noise (AWGN) channel impairments. Results reveal a pronounced asymmetry: semantic classification remains stable across a broad SNR range, even when reconstruction quality degrades significantly. These findings indicate that reliable task-level inference does not require high-fidelity geometric recovery and suggest that industrial wireless perception systems should prioritize semantic robustness under bandwidth and power constraints.

\section{Background and Related Work}

\subsection{Point Cloud Representation and Processing}

\subsubsection{Conventional Point Cloud Compression}

A 3D PC is a set of points in Euclidean space carrying geometric coordinates and optional attributes (e.g., color or reflectance). Due to their high dimensionality and density, efficient compression is essential for storage and transmission. Conventional point cloud compression (PCC) methods can be broadly categorized into three families~\cite{cao20193d}: 
(i) 1D traversal approaches, which serialize geometry into a one-dimensional representation prior to conventional compression; 
(ii) projection-based approaches, which map 3D data to 2D domains and leverage established image or video codecs; and 
(iii) volumetric approaches, which directly exploit spatial correlations using hierarchical spatial data structures such as octrees or K-dimensional (KD)-trees.

These principles are reflected in the MPEG standardization efforts, which define two major frameworks: Video-based PCC (V-PCC) and Geometry-based PCC (G-PCC). V-PCC projects dynamic PCs into 2D patches and compresses geometry and texture using conventional video codecs~\cite{schwarz2018emerging}. In contrast, G-PCC employs octree-based volumetric decomposition to encode geometric structure efficiently. Recent extensions enhance context modeling and rate–distortion optimization to improve compression performance~\cite{wang2024optimized}.

%While these approaches achieve high geometric compression efficiency, they treat point clouds primarily as geometric signals. As a result, they do not explicitly account for the semantic information required for downstream perception and decision-making tasks.

\subsubsection{Learning-Based Point Cloud Processing}

Deep learning has become the dominant paradigm for 3D perception tasks, including classification, segmentation, object detection, pose estimation, and reconstruction~\cite{guo2021deep, bello2020review}. However, unlike structured 2D images, PCs are irregular, unordered, and sparse, posing significant challenges for neural processing.

Three principal learning paradigms have emerged: 
(i) multi-view approaches that project 3D data into 2D representations processed by CNNs; 
(ii) voxel-based approaches that discretize space into regular volumetric grids; and 
(iii) point-based approaches that operate directly on raw point coordinates.

PointNet~\cite{qi2017pointnet} introduced a permutation-invariant architecture based on shared multi-layer perceptrons (MLPs) and symmetric pooling functions, establishing a foundation for direct point-based learning. However, its independent per-point processing limits local structural modeling. PointNet++~\cite{qi2017pointnet++} addresses this limitation via hierarchical neighborhood aggregation, enabling multi-scale geometric feature extraction. Graph-based models such as DGCNN~\cite{wang2019dynamic}  refine local context modeling through dynamic edge convolutions. More recently, transformer-based architectures~\cite{zhao2021point, wu2024pointv3} leverage self-attention mechanisms to capture long-range dependencies and global geometric relationships.

%In this work, PointNet++ is adopted due to its favorable balance between representational power and computational efficiency. Its hierarchical inductive bias is well-suited for capturing discriminative geometric structures in industrial point cloud data without the computational overhead of large-scale Transformer models.

\subsection{Semantic and Task-Oriented Communications}
\label{sec:semantic_comm}

Semantic and task-oriented communications redefine the objective of communication systems by shifting the focus from faithful signal reconstruction to reliable task execution. Rather than reproducing the source signal $\mathbf{X}$, the goal is to enable accurate estimation of a task-relevant output $T(\mathbf{X})$, such as a class label or control action. This paradigm, conceptually rooted in Weaver’s extension of Shannon’s framework~\cite{weaver_1953}, prioritizes task-relevant information while discarding redundant details, thereby improving communication efficiency~\cite{papas_2021}.

A formal foundation for this perspective is provided by the Information Bottleneck (IB) principle~\cite{tishby_2000}, which seeks a compressed representation $\mathbf{z}$ that preserves maximal information about the task output while minimizing redundancy with respect to the input:
\begin{equation}
    \min I(\mathbf{X}; \mathbf{z}) 
    \quad \text{subject to} \quad 
    I(\mathbf{z}; T(\mathbf{X})) \geq \epsilon,
\end{equation}
where $I(\cdot;\cdot)$ denotes mutual information and $\epsilon > 0$ is a predefined threshold that specifies the minimum amount of task-relevant information that must be preserved in the $\mathbf{z}$.% in the latent representation.

The semantic communication paradigm is particularly well aligned with IIoT scenarios where transmitted PCs primarily support downstream inference rather than geometric visualization. In such settings, the objective shifts from accurate coordinate reconstruction to reliable task execution under communication constraints. Task-aware representations therefore provide a natural foundation for aligning wireless transmission with decision-level robustness~\cite{Liu_2025}.

%\subsubsection{End-to-End Learned Communication Systems}

\textbf{End-to-End Learned Communication Systems:} End-to-end learned communication systems replace the traditional modular design, separating source coding, channel coding, and modulation, with jointly optimized neural transmitter–receiver architectures~\cite{oshea_2017}. In this formulation, the transmitter is modeled as an encoder $f_{\mathrm{enc}}(\cdot)$ that maps input data to channel symbols, which are corrupted by a differentiable stochastic channel model. The receiver, modeled as $f_{\mathrm{dec}}(\cdot)$, produces either a reconstructed signal or a task-specific output.

Joint optimization enables the learned representation to implicitly balance compression, robustness, and task performance. Unlike conventional pipelines with explicit channel coding, robustness to channel impairments emerges from the end-to-end training objective. The encoder learns to shape transmitted features in a manner that is resilient to stochastic channel distortions while preserving task-relevant information.

From a semantic communication perspective, the proposed formulation enables task-oriented transmission by directly optimizing inference performance rather than symbol-level fidelity. This is particularly relevant for high-dimensional 3D PC data, where conventional geometric compression does not necessarily preserve task-discriminative features. By jointly optimizing representation learning and transmission over the wireless channel, the framework aligns communication objectives with downstream perception tasks.

\section{System Model and Proposed Architecture}

\subsection{End-to-End Semantic Transmission Model}

\begin{figure}[t]
\centering
\includegraphics[width=1\linewidth]{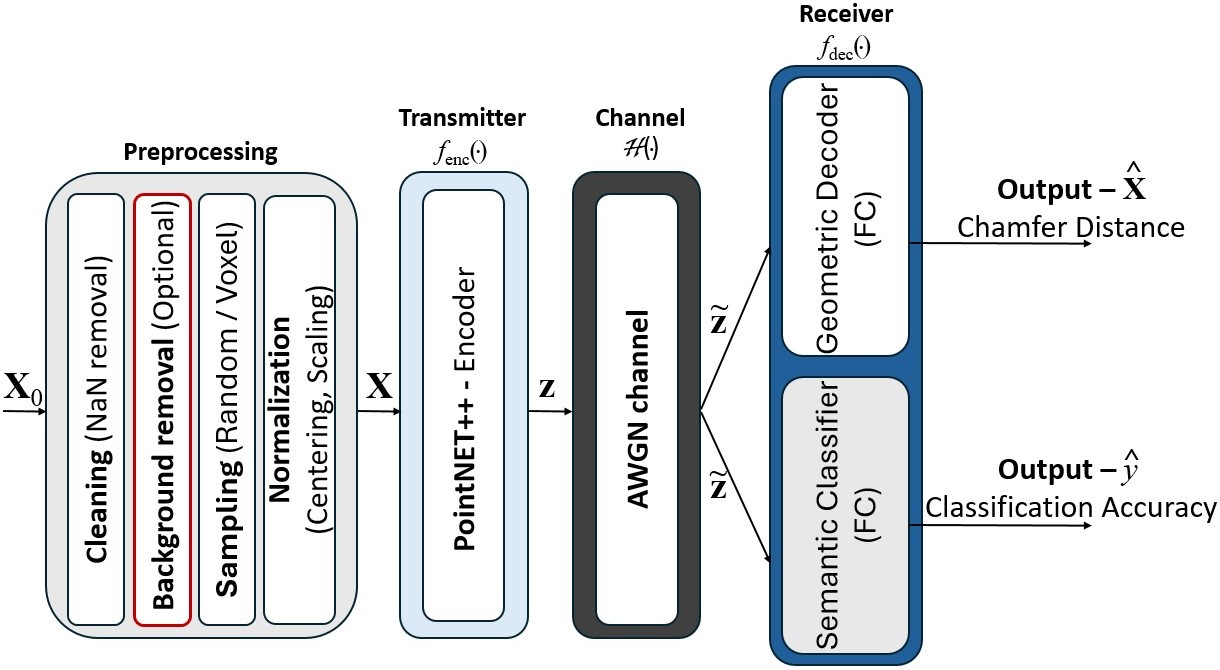}
\caption{End-to-end semantic PC transmission framework with parallel geometric reconstruction and semantic classification. The background removal step (red rectangle) is optional and evaluated separately in Section~IV.}
\label{fig:system_model}
\end{figure}

We consider an end-to-end learned semantic communication system for wireless transmission of 3D PCs in an industrial collaborative robotics scenario, as illustrated in Fig.~\ref{fig:system_model}. The system is designed to jointly support geometric reconstruction and semantic inference under wireless channel impairments. Formally, let $\mathbf{X}_0 \in \mathbb{R}^{N_0 \times 3}$ denote a raw PC acquired by a depth sensor, where $N_0$ may vary across samples. A deterministic preprocessing stage performs  sampling, and normalization, producing a fixed-size representation
$
\mathbf{X} \in \mathbb{R}^{N \times 3},
$
where $N$ is constant and compatible with the encoder architecture.

The preprocessed PC is mapped to a compact latent representation through a learnable encoder:
\begin{equation}
\mathbf{z} = f_{\mathrm{enc}}(\mathbf{X}; \boldsymbol{\theta}_{\mathrm{enc}}),
\end{equation}
where $\mathbf{z} \in \mathbb{R}^{D}$ and $D \ll 3N$. The latent vector $\mathbf{z}$ constitutes the transmitted representation and serves as the sole information carrier across the wireless channel. The communication channel is modeled as a stochastic transformation
$
\tilde{\mathbf{z}} = \mathcal{H}(\mathbf{z}),
$
where $\mathcal{H}(\cdot)$ captures channel impairments and is implemented as a differentiable layer during training.% This enables joint optimization of transmitter and receiver components.

At the receiver, the noisy latent representation $\tilde{\mathbf{z}}$ is processed by a task-aware decoder
$
\left( \hat{\mathbf{X}}, \hat{y} \right) 
= f_{\mathrm{dec}}(\tilde{\mathbf{z}}; \boldsymbol{\theta}_{\mathrm{dec}}),
$
where $f_{\mathrm{dec}}$ is implemented using two parallel task-specific branches. First, a geometric decoder reconstructs a PC estimate:
\begin{equation} \hat{\mathbf{X}} = f_{\mathrm{rec}}(\tilde{\mathbf{z}}; \boldsymbol{\theta}_{\mathrm{rec}}), 
\end{equation}
where $f_{\mathrm{rec}}(\cdot)$ aims to preserve the spatial structure of the original PC. Second, a semantic classifier produces a task-level prediction: 
\begin{equation} \hat{y} = f_{\mathrm{cls}}(\tilde{\mathbf{z}}; \boldsymbol{\theta}_{\mathrm{cls}}),
\end{equation}
where $\hat{y}$ corresponds to the object class relevant to the task. 

The encoder, channel model, and both decoder branches are trained jointly in an end-to-end manner by minimizing a composite loss function that combines reconstruction fidelity and semantic accuracy. This formulation enables the learned latent representation to simultaneously support geometric reconstruction and semantic reasoning while adapting implicitly to channel impairments. %Detailed descriptions of the encoder architecture, channel model, and receiver modules are provided in the following subsections.

\subsection{Transmitter Architecture \& Channel Model}

\subsubsection{PointNet++ Encoder} The transmitter is implemented using a PointNet++ encoder \cite{qi2017pointnet++}, a hierarchical architecture designed to learn compact and expressive representations of unordered point sets. Its multi-scale abstraction mechanism enables robust extraction of both local geometric structures and global shape information.  The encoder consists of multiple stacked Set Abstraction (SA) layers forming a hierarchical feature extraction pipeline. Each SA layer performs centroid selection via farthest point sampling (FPS), local neighborhood grouping using $k$-nearest neighbors (kNN), and shared feature extraction through a  MLP followed by symmetric max pooling to ensure permutation invariance.

Given an input point set $\mathbf{X} \in \mathbb{R}^{N \times (d+Ch)}$, where $d=3$ denotes spatial coordinates and $Ch$ represents feature channels from the previous abstraction level, an SA layer reduces the number of points to $N'$ and produces output features of dimension $Ch'$. Stacking SA layers progressively decreases spatial resolution while increasing feature abstraction.

The final abstraction stage aggregates all local features into a single global descriptor via symmetric max pooling, yielding a compact latent representation $\mathbf{z} \in \mathbb{R}^{D}$. Only this latent vector is transmitted over the wireless channel, enforcing strict transmitter–receiver separation. The sampling sizes, neighborhood parameters, MLP dimensions, and latent dimension $D$ are architectural hyperparameters whose specific values are provided in Section~IV-A1.

%\subsection{Wireless Channel Model}

\subsubsection{Wireless Channel Model} The wireless link is modeled as an AWGN channel acting directly on the latent representation. Prior to transmission, $\mathbf{z}$ is normalized to satisfy an average power constraint 
$\mathbb{E}\left[\|\mathbf{z}\|_2^2\right] = D.$ The received latent vector is defined as 
$
\tilde{\mathbf{z}} = \mathbf{z} + \mathbf{n},
$
where $\mathbf{n} \sim \mathcal{N}(\mathbf{0}, \sigma^2 \mathbf{I})$ and the variance $\sigma^2$ is determined by the signal-to-noise ratio (SNR),
$\mathrm{SNR} = 
\mathbb{E}\left[\|\mathbf{z}\|_2^2\right]/
\mathbb{E}\left[\|\mathbf{n}\|_2^2\right].$
During training, the SNR is uniformly sampled from a predefined range chosen to reflect practical short-range industrial wireless conditions. 

\subsection{Parallel Semantic and Geometric Decoders}

At the receiver, the noisy latent representation $\tilde{\mathbf{z}}$ is processed by two parallel modules, as illustrated in Fig. \ref{fig:system_model}, with distinct objectives: semantic inference and geometric reconstruction. This dual-decoder structure enables direct investigation of the relationship between semantic robustness and geometric fidelity under channel impairments.

\subsubsection{Geometric Decoder}
The geometric decoder reconstructs a fixed-size PC 
$\hat{\mathbf{X}} \in \mathbb{R}^{N \times 3}$
directly from $\tilde{\mathbf{z}}$ using a fully connected (FC) architecture. Hierarchical decoders based on PointNet++ feature propagation are commonly used in perception pipelines. However, such designs rely on intermediate encoder-side geometric structures and multi-scale feature hierarchies. In a communication setting, these intermediate representations are not inherently available at the receiver unless explicitly transmitted, which would violate strict transmitter–receiver separation.

The proposed FC decoder depends solely on the received latent vector. This enforces architectural consistency with the end-to-end communication model and guarantees that both semantic and geometric outputs originate from the same transmitted representation. This unified design enables a principled investigation of the trade-off between geometric fidelity and semantic reliability in wireless PC transmission.

\subsubsection{Semantic Decoder}
The semantic decoder maps $\tilde{\mathbf{z}}$ directly to a task-level prediction $\hat{y}$ using a lightweight FC network. Semantic inference primarily depends on discriminative global characteristics rather than precise point-level geometry. Consequently, operating on the compact latent representation is sufficient for reliable classification. The decoder relies exclusively on transmitted information, ensuring that all task-relevant content passes through the communication channel.

\begin{figure}
    \centering
    \includegraphics[width=0.9\columnwidth]{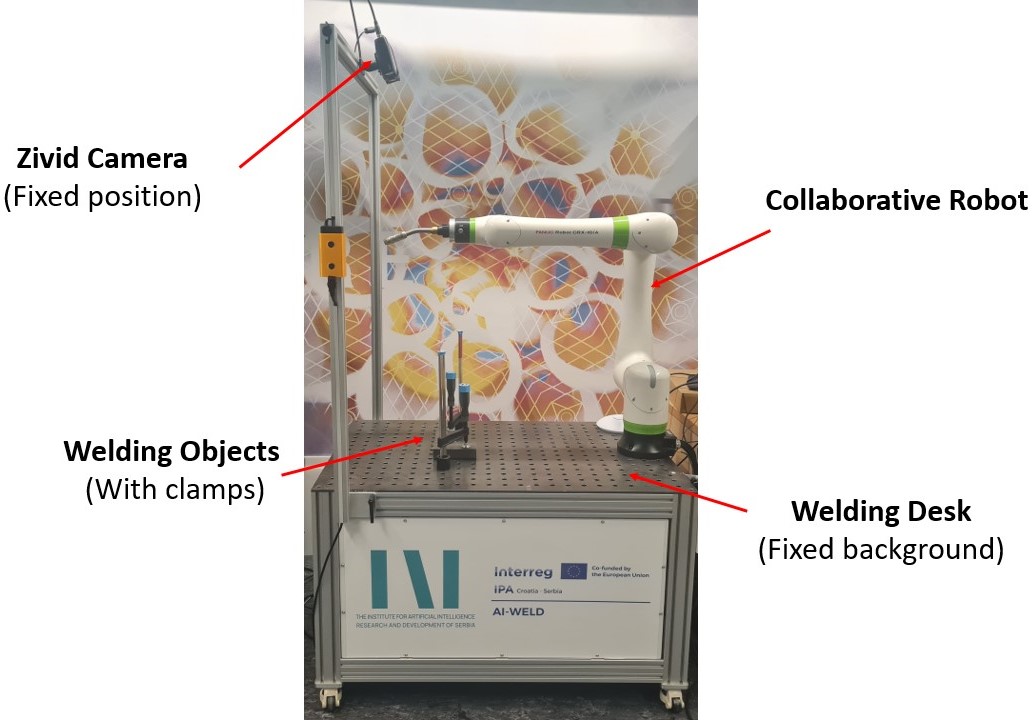}
    \caption{Industrial collaborative robotics data acquisition setup.}
    \label{fig:setup}
\end{figure}

\section{Performance Evaluation}

\subsection{Dataset and Experimental Setup}
To evaluate the proposed end-to-end semantic PC communication system under realistic industrial conditions, we construct a dedicated dataset acquired in a collaborative robotics laboratory. PCs are captured using a Zivid structured-light 3D camera rigidly mounted above a welding workbench, emulating a fixed overhead sensing configuration commonly used in industrial inspection and robotic manipulation. An overview of the setup is shown in Fig.~\ref{fig:setup}. During acquisition, objects are placed at random positions and orientations within the camera’s field of view. This introduces natural variations in pose, partial occlusions, and background clutter while maintaining consistent sensing geometry. The resulting dataset reflects practical industrial perception conditions.

The dataset comprises six object classes relevant to collaborative welding and assembly tasks. Four classes correspond to plastic components (two cylindrical objects, L-shaped pipes, T-shaped pipes, and L-shaped solid objects), while two classes correspond to metallic components (single-square and double-square objects). Representative sample is shown in Fig.~\ref{fig:placeholder} a).

For each class, 60 PCs are recorded, yielding a total of 360 samples. To increase scene variability, half of the samples (30 per class) are captured without welding clamps, and the remaining half include clamps present in the workspace. The inclusion of clamps introduces structured occlusions and additional geometric complexity, enabling evaluation of semantic robustness under challenging visual conditions.

\subsubsection{Training Procedure}
The proposed semantic PC communication framework is trained end-to-end in a supervised manner, with careful preprocessing and task-aware optimization to ensure stable convergence and robust semantic performance.

\begin{figure}[t]
    \centering
    \includegraphics[width=1\linewidth]{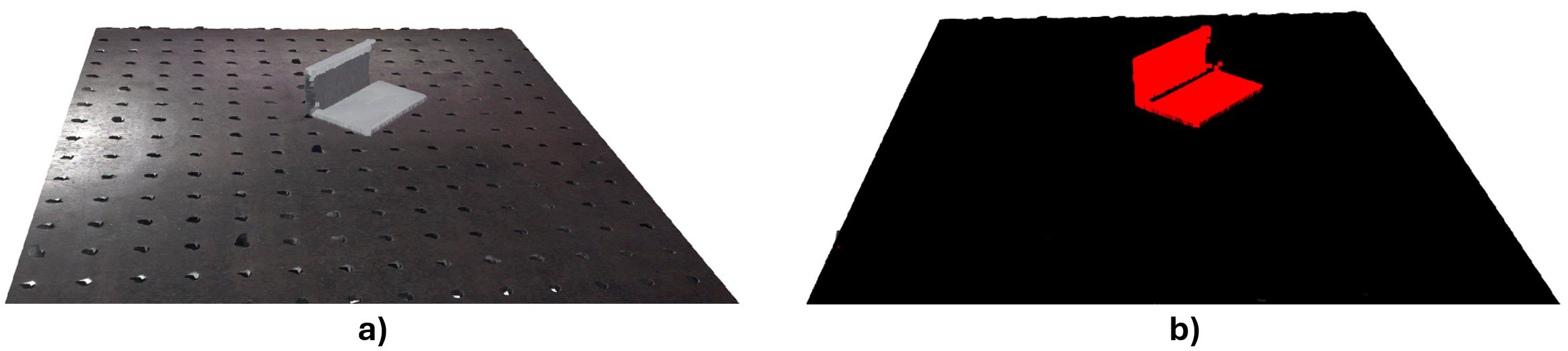}
    \caption{Representative dataset samples: a) Point cloud of an L-shaped solid object; b) Background removal prior to normalization and sampling.}
    \label{fig:placeholder}
\end{figure}

\textbf{Preprocessing:}
Each PC undergoes deterministic preprocessing to produce a fixed-size normalized input  (Fig.~\ref{fig:system_model}, leftmost part). Background removal is performed using a reference scan of the empty workspace to isolate object-centric geometry, as shown in Fig. \ref{fig:placeholder} b). Alternative configurations without background removal were also evaluated; however, they resulted in a disproportionate representation of background points during sampling and reduced semantic discriminability. Quantitative comparisons are provided in Sections~IV-B and IV-C.

The remaining object-centric points are resampled to $N = 2048$ points. If fewer than $N$ points are available, sampling with replacement is applied. Random sampling preserves the original point distribution, while voxel-based downsampling enforces spatial regularization through grid quantization. This allows evaluation of how different resampling strategies affect geometric reconstruction and semantic robustness. The PC is subsequently normalized by subtracting its centroid and scaling by the maximum Euclidean norm.

\textbf{Transmitter Architecture:}
For experimental evaluation, the transmitter employs a PointNet++ encoder with single-scale grouping instantiated with three SA layers. The first SA layer samples 512 centroids using FPS, groups 32 neighbors per centroid, and applies a shared MLP with output dimensions $(64, 64, 128)$. The second SA layer samples 128 centroids with 32 neighbors and applies an MLP of size $(128, 128, 256)$. The final SA layer performs global feature aggregation using an MLP of size $(256, 512, 1024)$ followed by symmetric max pooling, producing a 1024-dimensional global descriptor.
The global descriptor is processed by two FC layers of sizes 512 and $D$, generating the transmitted latent representation $\mathbf{z} \in \mathbb{R}^{D}$. To evaluate the impact of compression, we consider $D \in \{64, 256\}$. 

\textbf{Channel Modeling:}
Prior to transmission, the latent vector $\mathbf{z}$ is normalized using batch normalization to satisfy the average transmit power constraint. The normalized representation is then transmitted over an AWGN channel (see Section~III-B2). During training, the SNR is randomly sampled from $\{-4,-2,0,2,4,6,8\}$~dB on a per-batch basis to enforce robustness against channel variations and improve generalization to mismatched operating conditions.

\textbf{Receiver Architecture:}
At the receiver, the noisy latent vector $\tilde{\mathbf{z}}$ is processed by two parallel branches. The semantic decoder consists of FC layers of sizes $(D, 128, C)$ followed by softmax activation, corresponding to the $C$ object classes. The geometric decoder consists of FC layers of sizes $(D, 512, 1024, 6144)$, where the final layer outputs a reconstructed PC $\hat{\mathbf{X}} \in \mathbb{R}^{2048 \times 3}$. Both branches operate on the same received latent representation, enabling joint optimization of semantic and geometric objectives.

\textbf{Loss Function:}
The system is trained using a composite objective that combines geometric reconstruction and semantic classification losses. The reconstruction loss $\mathcal{L}_{\mathrm{rec}}$ is defined as the Chamfer distance (CD)\footnote{$\mathcal{L}_{\mathrm{rec}} = 
\sum_{\mathbf{x} \in \mathbf{X}} \min_{\hat{\mathbf{x}} \in \hat{\mathbf{X}}} \|\mathbf{x}-\hat{\mathbf{x}}\|_2^2
+
\sum_{\hat{\mathbf{x}} \in \hat{\mathbf{X}}} \min_{\mathbf{x} \in \mathbf{X}} \|\hat{\mathbf{x}}-\mathbf{x}\|_2^2$} ~\cite{NEURIPS2021_f3bd5ad5} between the original $\mathbf{X}$ and reconstructed $\hat{\mathbf{X}}$ PCs, while the classification loss $\mathcal{L}_{\mathrm{cls}}$ is defined as categorical cross-entropy. The overall objective is:
\begin{equation}
\label{total_loss}
\mathcal{L} = \mathcal{L}_{\mathrm{rec}} + \lambda_{\mathrm{cls}} \mathcal{L}_{\mathrm{cls}},
\end{equation}
where $\lambda_{\mathrm{cls}} \in \{0,1\}$ controls the inclusion of the semantic objective. Specifically, $\lambda_{\mathrm{cls}}=0$ corresponds to reconstruction-only training, while $\lambda_{\mathrm{cls}}=1$ enables joint geometric and semantic optimization. The impact of both configurations is evaluated in Sections~IV-B and IV-C.

\textbf{Optimization:}
Training is performed in an end-to-end manner using the Adam optimizer with learning rate $7\times10^{-4}$ and batch size $16$ for $50$ epochs. A $70/30$ train–test split with balanced class distribution is employed. After training, performance is evaluated under fixed SNR conditions to separately assess semantic accuracy and geometric reconstruction quality.

\subsection{Geometric Reconstruction Performance}
In this subsection, we evaluate the geometric reconstruction performance of the proposed framework over an AWGN channel and analyze the impact of key design choices. Reconstruction quality is quantified using the CD. To isolate purely geometric effects, reconstruction-only training ($\lambda_{\mathrm{cls}}=0$ in Eq~\eqref{total_loss}) is employed for the voxel-based sampling configuration, as voxelization primarily targets spatial uniformity rather than semantic discriminability. All other configurations, including random sampling with and without background removal, are trained jointly with $\lambda_{\mathrm{cls}}=1$ in Eq.~\eqref{total_loss}.

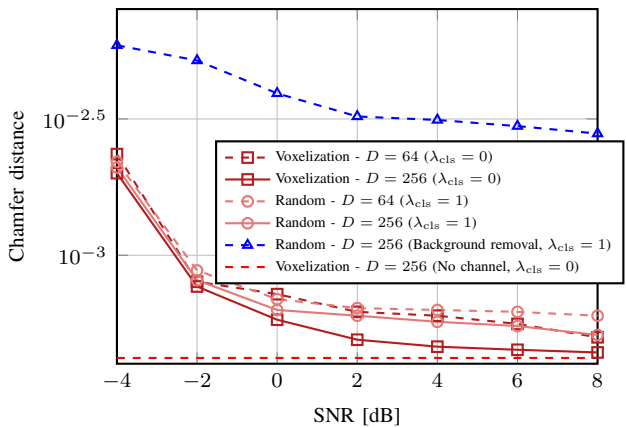
\begin{figure}[t]
\centering
\begin{tikzpicture}
  	\begin{semilogyaxis}[width=0.9\columnwidth, height=6.3cm, 
	legend style={at={(0.63,0.63)}, anchor= north,font=\scriptsize, legend style={nodes={scale=0.8, transform shape}}},
   	legend cell align={left},
	legend columns=1,   	 
   	x tick label style={/pgf/number format/.cd,
   	set thousands separator={},fixed},
   	y tick label style={/pgf/number format/.cd,fixed, precision=2, /tikz/.cd},
   	xlabel={SNR [dB]},
   	ylabel={Chamfer distance},
   	label style={font=\footnotesize},
   	grid=both,   	
   	xmin = -4, xmax = 8,
   	ymin=0.0004, ymax=0.008,
   	line width=0.8pt,
   	tick label style={font=\footnotesize},]
   	\addplot[recondark, dashed, mark=square, mark options=solid] 
   	table [x={x}, y={vox_64}] {./figs./reconstruction.txt};
   	\addlegendentry{Voxelization - $D=64$ ($\lambda_{\mathrm{cls}}=0$)}
    \addplot[recondark, mark=square] 
   	table [x={x}, y={vox_256}] {./figs./reconstruction.txt};
   	\addlegendentry{Voxelization - $D=256$ ($\lambda_{\mathrm{cls}}=0$)}

    	\addplot[reconlight, dashed, mark=o, mark options=solid] 
   	table [x={x}, y={rand_64}] {./figs./reconstruction.txt};
   	\addlegendentry{Random - $D=64$ ($\lambda_{\mathrm{cls}}=1$)}
    \addplot[reconlight, mark=o] 
   	table [x={x}, y={rand_256}] {./figs./reconstruction.txt};
   	\addlegendentry{Random - $D=256$ ($\lambda_{\mathrm{cls}}=1$)}

    	\addplot[blue, dashed, mark=triangle, mark options=solid] 
   	table [x={x}, y={rand_256_br}] {./figs./reconstruction.txt};
   	\addlegendentry{Random - $D=256$ (Background removal, $\lambda_{\mathrm{cls}}=1$)}
    \addplot[red, dashed] 
   	table [x={x}, y={no_channel}] {./figs./reconstruction.txt};
   	\addlegendentry{Voxelization - $D=256$ (No channel, $\lambda_{\mathrm{cls}}=0$)}

 	\end{semilogyaxis}
	\end{tikzpicture}
	\vspace*{-3mm}
\caption{Geometric reconstruction performance vs. SNR for different latent dimensions, sampling strategies, and background removal configurations.}
\label{Fig_rec}
\end{figure}

Fig.~\ref{Fig_rec} shows the reconstruction performance as a function of SNR for latent dimensions $D \in \{64, 256\}$. The system exhibits graceful degradation at low SNR and approaches the no-channel baseline with voxel-based sampling (red dashed line) at high SNR, indicating inherent robustness of the learned latent representation. Increasing the latent dimension from $D=64$ to $D=256$ consistently improves reconstruction performance across all SNR levels due to increased representational capacity.
Voxel-based sampling achieves lower CD compared to random sampling, particularly at moderate and high SNR values, owing to its more uniform spatial coverage.

When background removal is not applied, reconstruction performance is significantly higher and nearly matches the no-channel baseline at high SNR. This behavior is attributed to the strong geometric prior introduced by the static desk surface, as shown in Fig.~\ref{fig:placeholder}, which simplifies the reconstruction task. After removing the background, the CD increases by approximately one order of magnitude across all SNR values, reflecting a more challenging and realistic object-centric reconstruction scenario. Nevertheless, performance improves monotonically with increasing SNR in all configurations, confirming robustness under practical conditions.

\subsection{Semantic Performance Under Noisy Channel Conditions}

In contrast to geometric reconstruction, semantic recognition depends on the preservation of discriminative structural features rather than precise coordinate-level fidelity. Voxel-based downsampling aggregates points within local spatial cells and can be interpreted as a spatial quantization and low-pass filtering operation over the underlying geometric manifold. While this smoothing improves spatial uniformity and geometric stability, it suppresses high-frequency structural variations that are critical for semantic discrimination, particularly for small or thin components. Consequently, voxelization reduces the richness of local geometric descriptors available to the encoder. For this reason, all semantic experiments are conducted using random sampling, which preserves the original point distribution and maintains fine-grained geometric detail.

\subsubsection{Binary Welding Clamp Classification}

We first consider binary classification ($C=2$) of welding clamp presence. This task represents a relatively simple detection problem due to the distinctive geometric structure of the clamp. Fig.~\ref{Fig_class} (solid and dashed light blue curves) shows classification accuracy as a function of SNR. High accuracy is maintained even at low SNR values, and performance increases monotonically with SNR, reaching near-perfect accuracy at high SNR. Only minor differences are observed between experiments conducted with and without background removal, indicating that the semantic encoder effectively focuses on dominant object-specific features for this binary task. 

\subsubsection{Multi-Class Industrial Object Classification}

\begin{figure}[t]
\centering
\begin{tikzpicture}
  	\begin{axis}[width=0.9\columnwidth, height=6.3cm, 
	legend style={at={(0.75,0.25)}, anchor= north,font=\scriptsize, legend style={nodes={scale=0.8, transform shape}}},
   	legend cell align={left},
	legend columns=1,   	 
   	x tick label style={/pgf/number format/.cd,
   	set thousands separator={},fixed},
   	y tick label style={/pgf/number format/.cd,fixed, precision=2, /tikz/.cd},
   	xlabel={SNR [dB]},
   	ylabel={Accuracy [\%]},
   	label style={font=\footnotesize},
   	grid=both,   	
   	xmin = -4, xmax = 8,
   	ymin=52, ymax=100,
   	line width=0.8pt,
   	ytick={50, 60, 70, 80, 90, 100},
    yticklabels={$50$, $60$, $70$, $80$, $90$, $100$},
   	tick label style={font=\footnotesize},]
   	\addplot[classdark,dashed, mark=square, mark options=solid] 
   	table [x={x}, y={rand_nbr}] {./figs./class.txt};
   	\addlegendentry{Object classification - No background removal}
    \addplot[classdark, mark=square, mark options=solid] 
   	table [x={x}, y={rand_br}] {./figs./class.txt};
   	\addlegendentry{Object classification - Background removal}

        	\addplot[classlight, dashed, mark=triangle, mark options=solid] 
   	table [x={x}, y={clamp}] {./figs./class.txt};
   	\addlegendentry{Clamp detection - No background removal}
    \addplot[classlight, mark=triangle, mark options=solid] 
   	table [x={x}, y={clamp_br}] {./figs./class.txt};
   	\addlegendentry{Clamp detection - Background removal}
 	\end{axis}
	\end{tikzpicture}
	\vspace*{-3mm}
\caption{Semantic classification accuracy vs. SNR for binary clamp detection and six-class industrial object recognition ($\lambda_{\mathrm{cls}}=1$ in Eq. \eqref{total_loss} in all experiments).}
\label{Fig_class}
\end{figure}
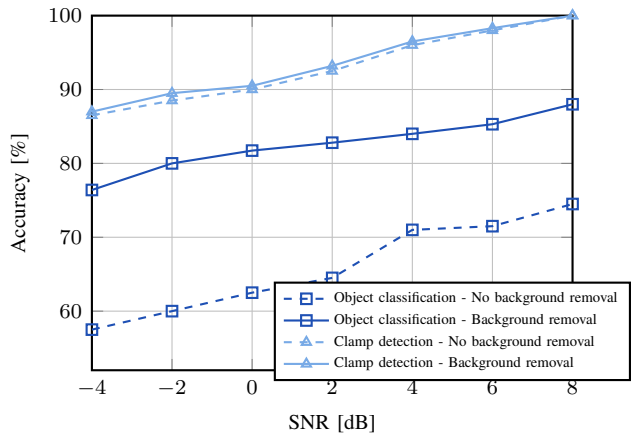

Fig.~\ref{Fig_class} (solid and dashed dark blue curves) presents classification accuracy versus SNR for six-class industrial object recognition ($C=6$). When background removal is applied, accuracy approaches $90\%$ at high SNR and degrades gradually as SNR decreases, indicating stable semantic robustness. In contrast, when random sampling is performed without prior background removal, accuracy saturates at approximately $70\%$, even at high SNR. This gap highlights the importance of background suppression in multi-class recognition. Without background removal, a substantial fraction of the $N=2048$ sampled points corresponds to the dominant planar desk surface, diluting object-specific features in the latent representation and reducing discriminative capacity.

%Compared to the binary task, overall accuracy is lower due to increased semantic complexity and partial geometric similarity among object classes. Nevertheless, achieving close to $90\%$ accuracy under realistic sensing conditions and noisy transmission confirms that the proposed semantic communication framework effectively preserves discriminative object features.

\subsection{Discussion: Semantic versus Geometric Robustness}

The experimental results highlight a clear distinction between geometric and semantic robustness under channel perturbations. Although both reconstruction and classification performance degrade as SNR decreases, their sensitivity to noise differs fundamentally. 

Geometric reconstruction, evaluated using the CD, is directly influenced by perturbations in the transmitted latent representation, as channel noise propagates to coordinate-level errors in the reconstructed PC. This sensitivity becomes more pronounced after background removal, where dominant planar structures no longer provide strong geometric priors. As a result, reconstruction quality reflects fine-grained spatial distortions and exhibits higher sensitivity to channel conditions.

In contrast, semantic performance demonstrates substantially greater resilience. Both binary welding clamp detection and six-class object classification maintain stable accuracy across a broad SNR range, with only gradual degradation at low SNR. This indicates that the learned latent representation preserves discriminative object-level features even when geometric precision is partially degraded. Small coordinate-level deviations that increase CD do not necessarily alter high-level semantic cues required for classification.

This asymmetry is particularly evident in the multi-class setting: despite a noticeable drop in reconstruction quality after background removal, semantic accuracy remains high, approaching $90\%$ at favorable SNR levels. These results confirm that reliable semantic inference does not require perfectly reconstructed geometry, but rather the preservation of structurally informative features within the latent space.

From a system-level perspective, this distinction has important implications for industrial wireless perception. In many practical scenarios, reliable decision-making is more critical than exact geometric reproduction. The results therefore suggest that end-to-end task-aware communication frameworks should prioritize semantic robustness over raw reconstruction fidelity under bandwidth and power constraints.

\section{Conclusion}

This paper presented an end-to-end semantic communication framework for wireless transmission of 3D PCs and provided a systematic comparison between geometric reconstruction fidelity and semantic robustness under channel impairments. Experimental results revealed a clear asymmetry: while geometric reconstruction degrades significantly at low SNR, semantic classification remains stable across a broad operating range, demonstrating that reliable task-level inference does not require high-fidelity geometric recovery. These findings suggest that future industrial wireless perception systems should prioritize semantic robustness over strict coordinate-level accuracy under bandwidth and power constraints. Future work will focus on real-world wireless implementation of the proposed framework and comprehensive system-level analysis including latency, bandwidth efficiency, and energy consumption.

%\begin{thebibliography}{00}
%\bibitem{b1} G. Eason, B. Noble, and I. N. Sneddon, ``On certain integrals of Lipschitz-Hankel type involving products of Bessel functions,'' Phil. Trans. Roy. Soc. London, vol. A247, pp. 529--551, April 1955.
%\end{thebibliography}

\bibliographystyle{IEEEtran}
\bibliography{references}

@inproceedings{cao20193d,
  title={3D point cloud compression: A survey},
  author={Cao, Chao and Preda, Marius and Zaharia, Titus},
  booktitle={Proc. 24th Int. Conf. 3D Web Technol.},
  pages={1--9},
  year={2019}
}

@article{schwarz2018emerging,
  title={Emerging MPEG standards for point cloud compression},
  author={Schwarz, Sebastian and others},
  journal={IEEE J. Emerg. Sel. Topics Circuits Syst.},
  volume={9},
  number={1},
  pages={133--148},
  year={2018},
  publisher={IEEE}
}

@article{wang2024optimized,
  title={Optimized octree codec for geometry-based point cloud compression},
  author={Wang, Zhecheng and Wan, Shuai and Wei, Lei},
  journal={Signal, Image and Video Process.},
  volume={18},
  number={1},
  pages={761--772},
  year={2024},
  publisher={Springer}
}

@Article{bello2020review,
    AUTHOR = {Bello, Saifullahi Aminu and Yu, Shangshu and Wang, Cheng and Adam, Jibril Muhmmad and Li, Jonathan},
    TITLE = {Review: Deep Learning on 3D Point Clouds},
    JOURNAL = {Remote Sens.},
    VOLUME = {12},
    YEAR = {2020},
    NUMBER = {11},
}

@ARTICLE{guo2021deep,
  author={Guo, Yulan and others},
  journal={IEEE Trans. Pattern Anal. Mach. Intell.}, 
  title={Deep Learning for 3D Point Clouds: A Survey}, 
  year={2021},
  volume={43},
  number={12},
  pages={4338-4364},
  doi={10.1109/TPAMI.2020.3005434}}

@inproceedings{qi2017pointnet,
  title={Pointnet: Deep learning on point sets for 3d classification and segmentation},
  author={Qi, Charles R and Su, Hao and Mo, Kaichun and Guibas, Leonidas J},
  booktitle={Proc. IEEE Conf. Comput. Vis. Pattern Recognit. (CVPR)},
  pages={652--660},
  year={2017}
}

@article{qi2017pointnet++,
  title={Pointnet++: Deep hierarchical feature learning on point sets in a metric space},
  author={Qi, Charles Ruizhongtai and Yi, Li and Su, Hao and Guibas, Leonidas J},
  journal={Adv. Neural Inf. Process. Syst.},
  volume={30},
  year={2017}
}

@article{wang2019dynamic,
  title={Dynamic graph cnn for learning on point clouds},
  author={Wang, Yue and others},
  journal={ACM Trans. Graph.},
  volume={38},
  number={5},
  pages={1--12},
  year={2019},
  publisher={Acm New York, NY, USA}
}

@inproceedings{zhao2021point,
    author    = {Zhao, Hengshuang and Jiang, Li and Jia, Jiaya and Torr, Philip H.S. and Koltun, Vladlen},
    title     = {Point Transformer},
    booktitle = {Proc. IEEE/CVF Int. Conf. Comput. Vis. (ICCV)},
    month     = {October},
    year      = {2021},
    pages     = {16259-16268}
}

@inproceedings{wu2024pointv3,
    author    = {Wu, Xiaoyang and others},
    title     = {Point Transformer V3: Simpler Faster Stronger},
    booktitle = {Proc. IEEE Conf. Comput. Vis. Pattern Recognit. (CVPR)},
    month     = {June},
    year      = {2024},
    pages     = {4840-4851}
}

@article{weaver_1953,
  title={Recent contributions to the mathematical theory of communication},
  author={W. Weaver},
  pages={261-281},
  journal={ETC: Rev. Gen. Semantics},
  volume={},
  year={1953}
}

@article{papas_2021,
  title={Semantics-empowered communication for networked intelligent systems},
  author={Kountouris, M. and Pappas, N.},
  pages={96-102},
  journal={IEEE Commun. Mag.},
  volume={59},
  number={6},
  year={2021}
}

@article{tishby_2000,
  title={The information bottleneck method},
  author={Tishby, N. and Pereira, F. C and Bialek, W.},
  pages={},
  journal={arXiv preprint, arXiv:physics/0004057},
  volume={},
  number={},
  year={2000}
}

@article{OShea_2017,
  title={An introduction to deep learning for the physical layer},
  author={O’Shea, T. and Hoydis, J.},
  pages={563-575},
  journal={IEEE Trans. Cogn. Commun. Netw.},
  volume={3},
  number={4},
  year={2017}
}

@article{Liu_2025,
  title={A Semantic Communication System for Point Cloud},
  author={Liu, X. and Liang, H. and Bao, Z. and Dong, C. and Xu, X.},
  pages={894-910},
  journal={IEEE Trans. Veh. Technol.},
  volume={74},
  number={1},
  year={2025}
}

@article{Shao_2025,
  title={Point Cloud in the Air},
  author={Shao, Y. and Bian, C. and Yang, L. and Zhang, Z. and Gündüz, D.},
  pages={142-148},
  journal={IEEE Commun. Mag},
  volume={63},
  number={12},
  year={2025}
}

@article{liu_2024,
  author    = {Yi Liu and Changsheng Zhang and Xingjun Dong and Jiaxu Ning},
  title     = {Point cloud-based deep learning in industrial production: A survey},
  journal   = {ACM Comput. Surv.},
  volume    = {57},
  number    = {7},
  pages     = {1--36},
  year      = {2024}
}

@ARTICLE{saleh_2025,
  author={S. Safarnejad and M. H. Soheilian and B. Safaei},
  journal={IEEE Internet Things J.},
  title={NORRIS: Noise-Resilient and Resource-Efficient Semantic Encoded Point Cloud Data Transmission for Internet of Things Communications},
  year={2025},
  volume={12},
  number={13},
  pages={25720--25731},
  doi={10.1109/JIOT.2025.3559435}
}

@ARTICLE{11361156,
  author={W. Yang and others},
  journal={IEEE Trans. Cogn. Commun. Netw.},
  title={Channel-Adaptive Cross-Modal Generative Semantic Communication for Point Cloud Transmission},
  year={2026},
  volume={12},
  pages={5983--5998},
  doi={10.1109/TCCN.2026.3657061}
}

@INPROCEEDINGS{10437861,
  author={T. Han and K. Chi and Q. Yang and Z. Shi},
  booktitle={Proc. IEEE Global Commun. Conf. (GLOBECOM)},
  title={Semantic-Aware Transmission for Robust Point Cloud Classification},
  year={2023},
  pages={7617--7622},
  doi={10.1109/GLOBECOM54140.2023.10437861}
}

@article{NEURIPS2021_f3bd5ad5,
 author = {Wu, Tong and others},
 journal = {Adv. Neural Inf. Process. Syst.},
 pages = {29088--29100},
 title = {Balanced Chamfer Distance as a Comprehensive Metric for Point Cloud Completion},
 volume = {34},
 year = {2021}
}

\end{document}